\documentclass[sn-mathphys,Numbered]{sn-jnl}

\usepackage{graphicx}%
\usepackage{multirow}%
\usepackage{amsmath,amssymb,amsfonts}%
\usepackage{amsthm}%
\usepackage{mathrsfs}%
\usepackage[title]{appendix}%
\usepackage{xcolor}%
\usepackage{textcomp}%
\usepackage{manyfoot}%
\usepackage{booktabs}%
\usepackage{algorithm}%
\usepackage{algorithmicx}%
\usepackage{algpseudocode}%
\usepackage{listings}%
\usepackage{caption}
\usepackage{subcaption}
\usepackage{color,soul}

\theoremstyle{thmstyleone}%

\theoremstyle{thmstyletwo}%

\theoremstyle{thmstylethree}%

\begin{document}

\title[Article Title]{Metasurface lens that is converging or diverging depending on transmission direction enables ultra-compact MEMS tunable reflective lens.}

\author[1]{\fnm{Firehun} \sur{Dullo}}\email{firehun.t.dullo@sintef.no}
\author[1]{\fnm{Jesil} \sur{Jose}}\email{jesil.jose@sintef.no}

\author[1]{\fnm{Gregory} \sur{Bouquet}}\email{gregory.bouquet@sintef.no}
\author[1]{\fnm{Zeljko} \sur{Skokic}}\email{zeljko.skokic@sintef.no}

\author*[1]{\fnm{Christopher} \sur{Dirdal}}\email{christopher.dirdal@sintef.no}

\affil*[1]{\orgdiv{Smart Sensors and Microsystems}, \orgname{SINTEF Digital}, \orgaddress{\street{Forskningsveien 1}, \city{Oslo}, \postcode{0373}, \country{Norway}}}

\abstract{A conventional refractive lens surface can act as a positive (converging) or negative (diverging) lens, but the same surface cannot act as both. We show that a geometric phase metasurface lens can have the unique property of acting both as a positive or negative lens upon transmission through its front or rear side, respectively. This offers certain freedom in compound lens design, where one combines focusing and defocusing operations. We utilize this property to make an ultra-compact, varifocal reflective lens, where a metasurface is placed in front of a novel long-stroke piezoelectric MEMS-micromirror. A large theoretical diopter tunability of 6330 m$^{-1}$ is enabled due to innovative thin-film piezoelectric MEMS design, offering 62 µm displacement at 40V and low power, along with rapid actuation in the kHz region. The achieved MEMS-displacement is an order of magnitude larger than  previously reported out-of-plane mechanical metasurface actuation. Since both metasurface and micromirror are flat, the presented reflective lens can be assembled without need for a spacer. It is therefore well suited for wafer-level silicon fabrication at high volumes and low cost. A proof-of-concept implementation using a 1550nm NIR metalens is demonstrated, attaining on the order of 1121 m$^{-1}$ diopter change for a focal length shift of 270 µm caused by a 53 µm micromirror displacement.}


\maketitle

\section{Introduction}\label{sec1}
\begin{figure}[h!]
    \centering
    \includegraphics[width=1.0\textwidth]{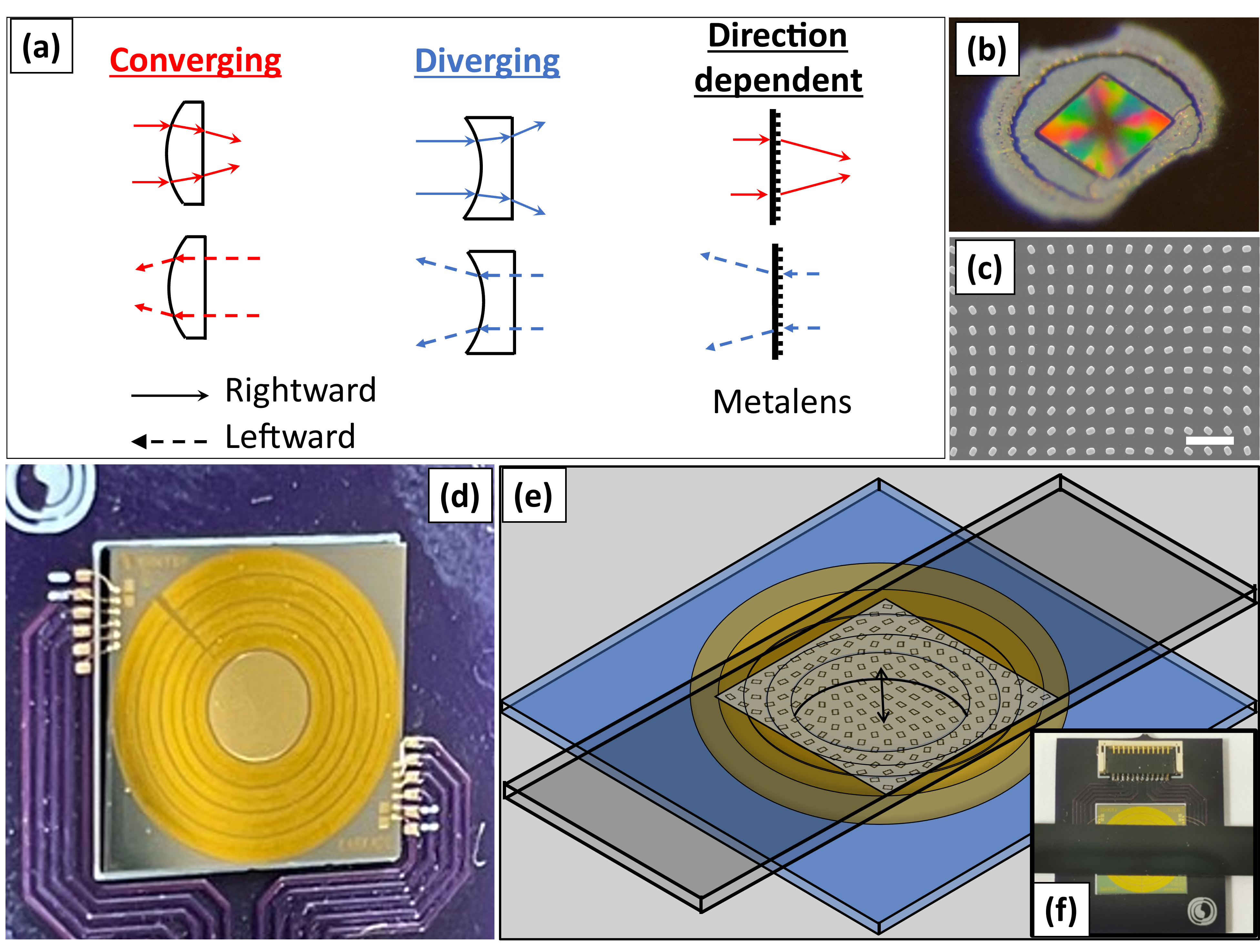}
    \caption{a) Conventional lenses either focus light or defocus light, whether the incoming light is rightward or leftward through them. Contrary to this, a geometric phase metasurface lens can enable focusing in one direction and defocusing in the other. (b) A NIR geometric phase metasurface lens (300µm x 300µm) and (c) a SEM image of a portion of the metasurface lens, where the scalebar represents 2µm. (d) A novel piezoelectric MEMS micromirror (die size 9 mm x 9 mm) capable of lifting the central mirror aperture up or down in a piston motion. (e) An illustration of the compact reflective metalens (CRM) concept where the dual converging/diverging metasurface on a silicon substrate is mounted on top of the MEMS micromirror. As the central mirror is drawn away from the metasurface, the focal length of the CRM is modulated. (f) An image of the assembled metasurface and MEMS mirror (the metasurface is faced down towards the mirror).}
    \label{fig:IllustrationImage}
\end{figure}

The earliest known use of lenses dates back several thousands of years to ancient Egypt \cite{enoch2007archeological}. A fairly modern concept of lenses arose already in 10th century AD when they were understood by the Persian mathematician Ibn Sahl in terms of refraction \cite{rashed1990pioneer}, and in 1609 Galileo presented an improved telescope which used convergent and divergent lenses that are easily recognizable as lenses today \cite{dupr2002galileo}. Throughout this long history, lenses have been either focusing or defocusing, meaning that if e.g. a lens focuses light passing through its front side, it will also focus light passing through its rear side (and vice versa). From common experience, turning a magnifying glass around does not change its optical function. This is because the property of focusing or defocusing corresponds to the convex or concave geometry of a lens, respectively (illustrated in Fig. \ref{fig:IllustrationImage}a), and this property is kept for both directions of light propagation. In this paper we demonstrate a notable possibility of breaking this rule. One can make static surfaces that are both focusing and defocusing depending on whether light propagates through their front or back side, unlike previous lenses. We show that this can be achieved using the recently developed geometric phase (also known as Pancharatnam-Berry phase) metasurface lenses \cite{khorasaninejad2016metalenses} under the condition that its sub-units are reflection symmetric and that they are placed according to the circular symmetry of the lens phase function. We show that particular design freedom is thereby offered for compound lens design, when one wishes to combine focusing and defocusing operations.

Until recently it has also been challenging to make compact lenses that can tune their focusing ability. This is a highly desirable property in a wide variety of optical systems. Changing the focal point and optical power (measured in diopters) of a lens enables functionalities like projecting at different distances, imaging at different depths or modulating the power incident on an area. Traditional tunable optical lenses typically rely on motors to displace lenses, leading to constraints in terms of size, cost, power consumption and response time. However, recent developments in micro-optical tunable lenses have paved a way for overcoming these limitations \cite {Kim2022-tunablemetalesreview}.  

Micro-optics offer fast, small, lightweight, and cost-effective tunability of relevance to miniaturized optical tunable lenses. These are properties necessary for a wide range of applications from biological imaging to applications on drones or robots \cite{Kim2022-tunablemetalesreview, jablonowski2020beyond}. There exist micro-optical techniques by which a single lens can be tuned (e.g. by deformation of polymer lenses \cite{kamali2016highly, She2018-memspolymerlens} or modulating the refractive index in e.g. liquid crystal lenses \cite{bosch2021-LCMSlens, zhong2016-LCMSlens}). However, for faster actuation rates and stronger modulations, MEMS-displaced lens doublets have emerged as a promising platform \cite{He2019-tunMEMSmsLensreview}: These include both modulating the gaps between the lenses \cite{arbabi2018mems, Dirdal2022mems} and sideways translation of lenses of the Alvarez lens type \cite{han2020mems, han2022mems}.

While traditional refractive lenses rely on curvature to achieve a specific focal length, flat lenses utilizing subwavelength nanostructures have recently emerged, known as metasurface lenses (metalenses). Whereas refractive lenses are typically bulky and add weight, metalenses allow for a compact form factor. These can furthermore be fabricated in silicon MEMS and CMOS processing lines \cite{dirdal2023uv}. In addition to the benefits in form factor and manufacturability, metasurface elements have been extensively explored for their added degrees of freedom in manipulating light \cite {Kim2022-tunablemetalesreview}, also being integrated in MEMS \cite {Xu2022-MEMSmetasurfaceReview, He2019-tunMEMSmsLensreview} or other tunability schemes \cite{Yoo2023-tunableMSreview, kang2019recent}. For MEMS based devices, the optical tunability is often constrained by the available MEMS displacement range. On that note, thin-film piezoelectric MEMS (piezoMEMS) has been proposed due to its ability to offer long-stroke quasi-static displacement at low voltages. Thin-film piezoMEMS was first applied to metasurfaces in our earlier work  \cite{meng2021dynamic}. In another study, we presented a varifocal doublet lens utilizing piezoMEMS, showcasing a maximum metalens displacement of 7.2 µm \cite{Dirdal2022mems}. Although this demonstrated a significant advancement, there is still room for improving both the displacement range and compactness of the system. 

Having a lens which can both focus and defocus depending on transmission direction gives certain design freedoms not available from conventional lenses. For instance, this enables making a reflective equivalent to Galileo's telescope by placing such a lens in front of a mirror. Rather than make a telescope, we combine this arrangement with long stroke piezoMEMS actuation to enable an ultra-compact reflective lens doublet. Our resulting device offers strong optical modulations and a diopter change on the order of 6330 m$^{-1}$ by displacing a piezoMEMS micromirror relative to a geometric phase metasurface lens by 62 $\mu$m  with 40V application. The concept and involved components are shown in Fig. \ref{fig:IllustrationImage}. The achieved MEMS-displacement range is more than an order of magnitude larger than state of the art electrostatic actuation of a metasurface doublet at half the voltage \cite{arbabi2018mems}, and an order of magnitude larger than in our previous work \cite{Dirdal2022mems}. The lens can be operated up to kHz range and is inherently low power due to utilizing thin-film piezoelectric PZT actuators. An ultra-compact form factor is enabled by a unique feature of using a geometric phase metalens: Notably, the metalens acts as a positive (converging) lens in one direction and a negative (diverging) lens in the other. Furthermore, the reflective arrangement of the device additionally doubles the effective mirror displacement compared to transmissive varifocal lenses. In combination these properties enable an ultra-compact lens which offers strong modulations.

\section{Theory} \label{Sec:Theory}

\subsection{Reflective doublets}
\begin{figure}[h!]
     \centering
     \begin{subfigure}[b]{0.7\textwidth}
         \centering
         \includegraphics[width=\textwidth]{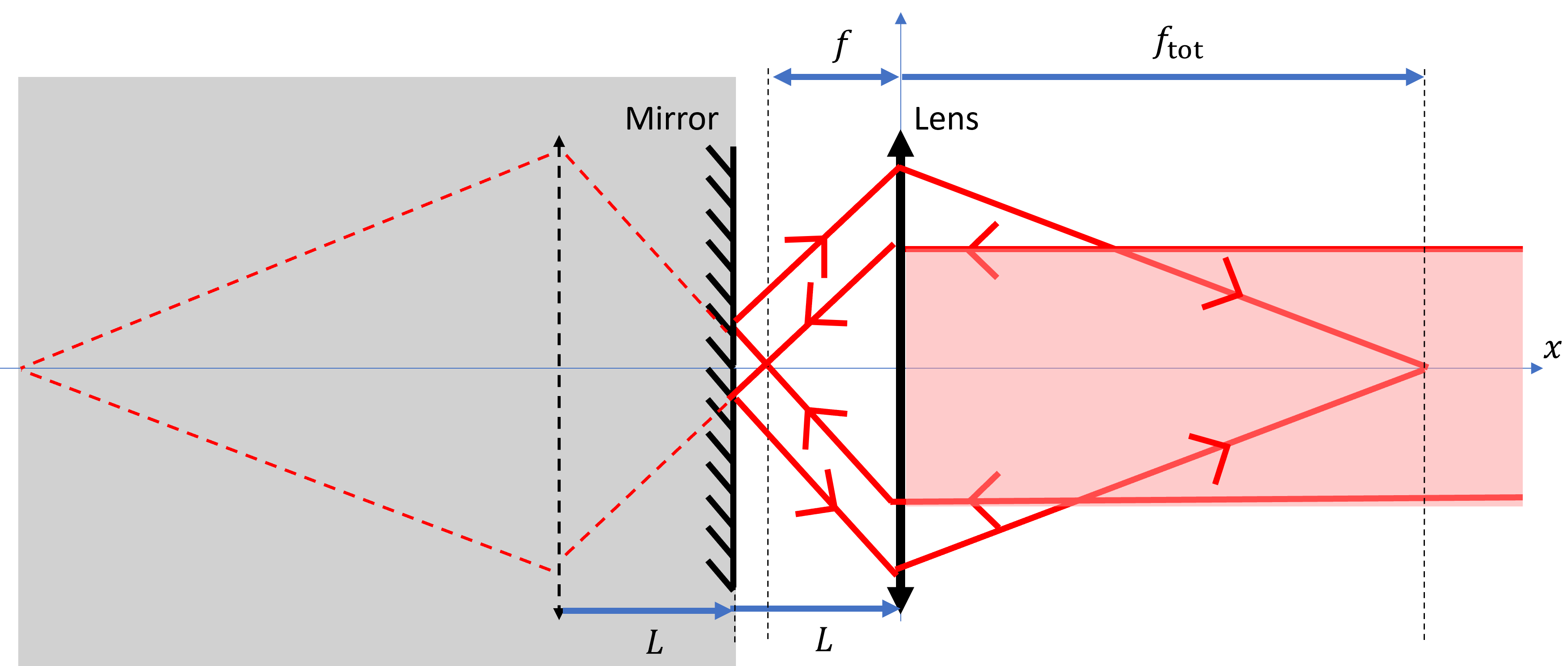}
         \caption{Reflective lens}
         \label{fig:ReflectiveLens}
     \end{subfigure}
     \hfill \\
     \begin{subfigure}[b]{0.45\textwidth}
         \centering
         \includegraphics[width=\textwidth]{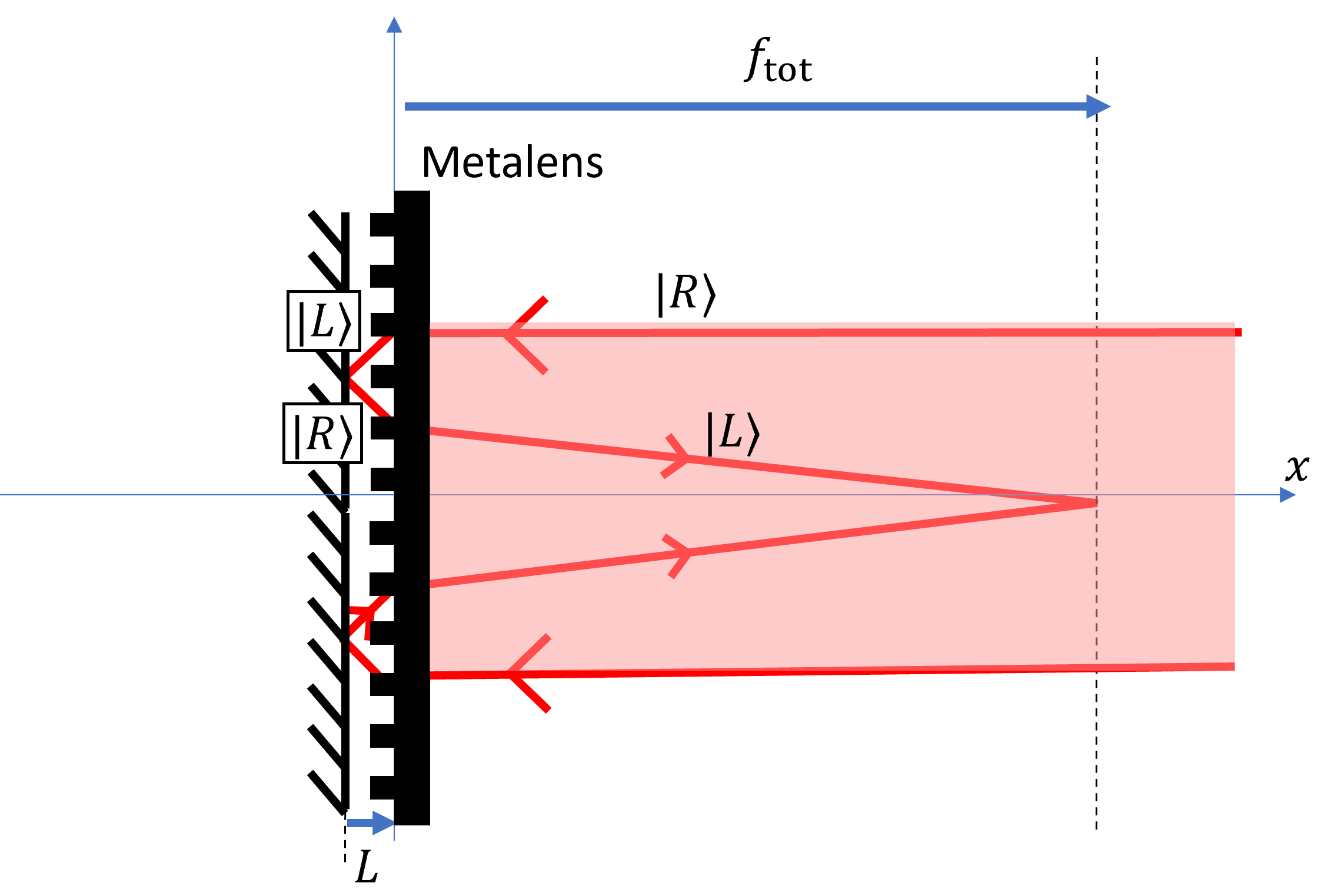}
         \caption{Compact reflective metalens (CRM)}
         \label{fig:MetaReflectiveLens}
     \end{subfigure}
        \caption{\textbf{Two implementations of a reflective lens.} (a) A reflective arrangement using a conventional lens combined with a mirror. The mirrored system is also drawn (within the gray box) to show that this reflective arrangement is equivalent to a transmissive arrangement of twice the lens-mirror spacing $2L$. A reflective arrangement is therefore useful when the mirror actuation distance is limited. When using a conventional refractive lens the lens-mirror gap $L$ must correspond to roughly the focal length $L\sim f$ in order to attain large focal length change upon displacing the mirror. With limited mirror displacement range, such an implementation may necessitate placing a spacer between the mirror and the lens, which could complicate wafer-level fabrication. (b) Using a geometric phase metasurface lens (metalens) gives design freedom which allows the lens-mirror spacing to be set to zero $L\to 0$ while maintaining large focal length changes. This is because the metalens acts as a positive (converging) lens to the incoming path and a negative (diverging) lens to the reflected path, enabling a reflective doublet effectively consisting of a negative and positive lens.}
        \label{fig:TwoReflectiveLenses}
\end{figure}

Figure \ref{fig:ReflectiveLens} shows the sketch of a reflective tunable lens consisting of a static lens with focal length $f$ placed in front of a mirror. The lens-mirror separation is given by $L$. The mirrored (virtual) system is also drawn within the gray box, revealing that this reflective arrangement is equivalent to a transmissive doublet of twice the lens-mirror separation $2L$. By modulating the separation distance, the incoming collimated light leaves the reflected lens as either focused, defocused or collimated. In micro-optic implementations of such a device, a MEMS mirror may be used for modulating the distance $L$, where available modulation ranges are typically restricted to some microns for electrostatic actuation \cite{arbabi2018mems} up to hundreds of microns for thermal actuation \cite{zhou2020mems}. The twice effective gap $2\Delta L$ of a reflective arrangement relative to a transmissive lens doublet is therefore particularly suited for MEMS-actuation where the lens-mirror displacement range is constrained. In Sec. \ref{Sec:MEMS} we present a thin-film piezoelectric MEMS architecture which offers lens-mirror displacements an order of magnitude larger than the state-of-the art electrostatic and piezoelectric displacement ranges.

In Fig. \ref{fig:ReflectiveLens} collimation occurs when $L=f$, i.e. when the incoming collimated light is focused to and reflected from the focal point of the lens. This is the lens-mirror separation at which the largest changes in focal length occur: Moving the focal point from infinite to finite distances. Therefore, a separation on the order of $L\sim f$ is required for a tunable lens which can offer large focal length changes. This is however a considerable gap for MEMS-mirror actuation. If focal lengths are desired in the $f\sim300$ µm range and upward the necessary gap $L\sim 300\mu$m is already on the length scale of a silicon wafer, and considerably supersedes typical MEMS-piston motion ranges. Fabrication of a MEMS-implementation of such a lens could thus require the addition of a spacer element between the MEMS and lens elements, which thus may complicate wafer-level fabrication.

\subsection{Geometric phase enables compact reflective metalens (CRM)} \label{sec:GPM_makes_CRM}
We shall now present an alternative design which makes use of unique properties of a geometric phase metasurface lens (GP-metalens). Fig. \ref{fig:MetaReflectiveLens} shows a similar arrangement for a tunable reflective lens, but this time a GP-metalens is placed in front of the mirror. It will be shown that the GP-metalens acts as a positive (converging) lens upon first transmission, and a negative (diverging) lens upon second transmission, after being reflected by the mirror. While the previous arrangement (Fig. \ref{fig:ReflectiveLens}) is equivalent to a transmissive doublet consisting of two equal positive lenses, the metasurface variant in Fig. \ref{fig:MetaReflectiveLens} is therefore equivalent with a transmissive doublet of a positive and negative lens of equal focal length magnitude, for which the largest changes in focal length occur near to zero lens-mirror gap $L \to 0$. This removes the need for a spacer between lens and mirror, allowing for the reflective lens to be ultra-compact and easily fabricated in wafer-level processes.

\begin{figure}[h]
     \centering
     \begin{subfigure}[b]{0.45\textwidth}
         \centering
         \includegraphics[width=\textwidth]{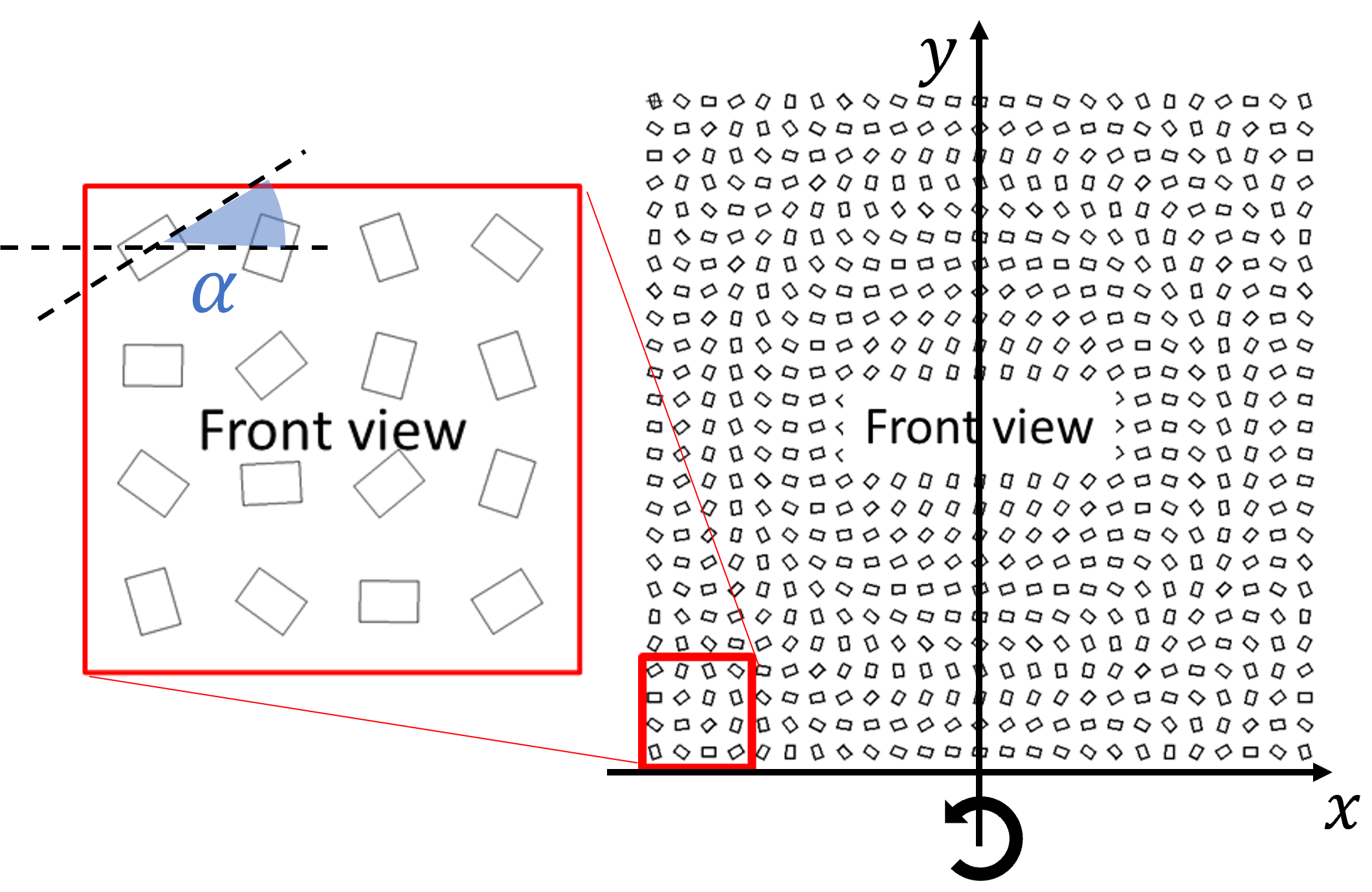}
         \caption{Front view.}
         \label{fig:IncomingMetasurface}
     \end{subfigure}
     \hfill 
     \begin{subfigure}[b]{0.45\textwidth}
         \centering
         \includegraphics[width=\textwidth]{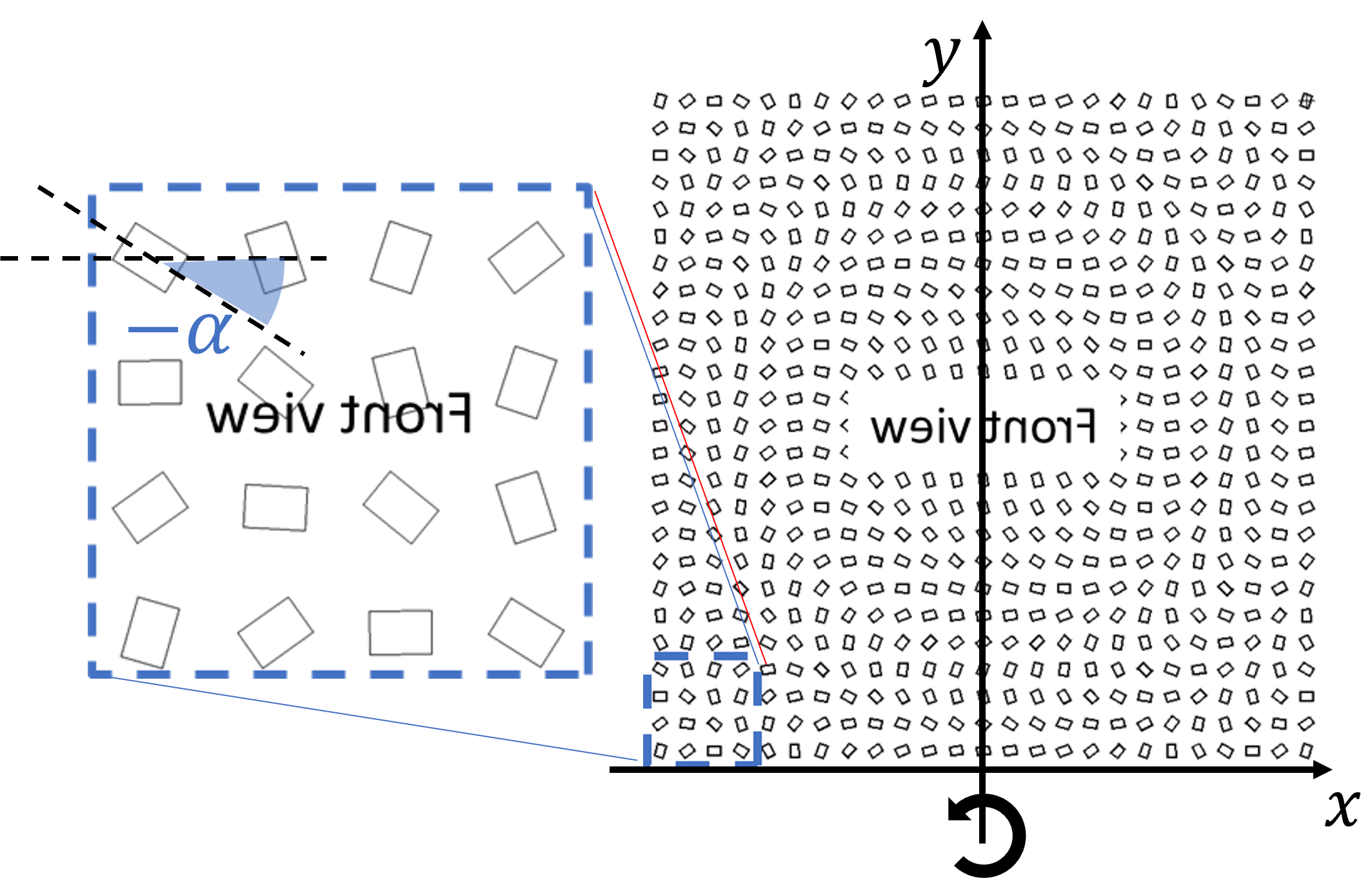}
         \caption{Rear view (reflected metasurface).}
         \label{fig:OutgoingMetasurface}
     \end{subfigure}
     \\ 
     \begin{subfigure}[b]{0.55\textwidth}
         \centering
         \includegraphics[width=\textwidth]{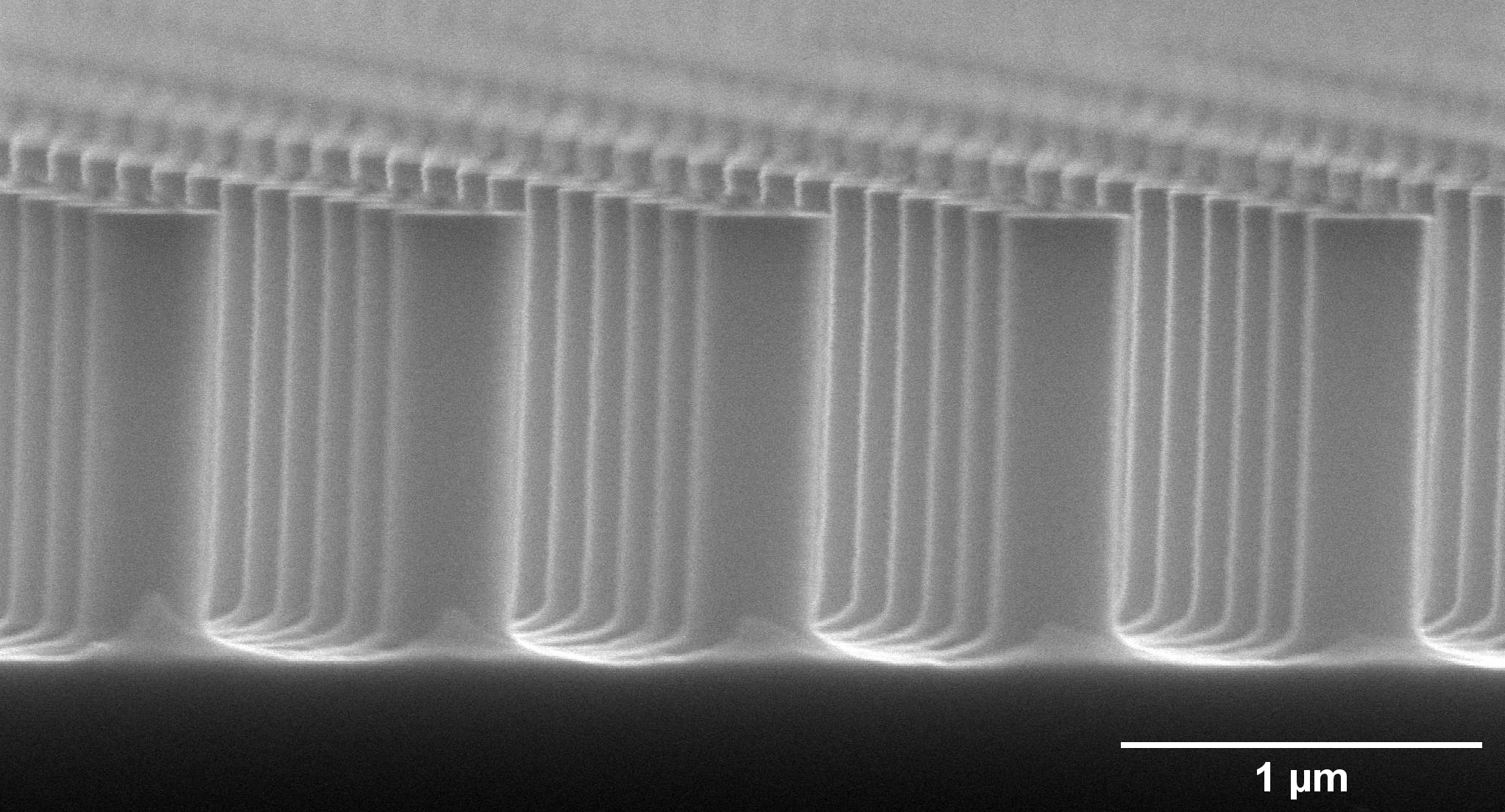}
         \caption{Cross-section}
         \label{fig:SEM_Crossection}
     \end{subfigure}
     \hfill 
     \begin{subfigure}[b]{0.37\textwidth}
         \centering
         \includegraphics[width=\textwidth]{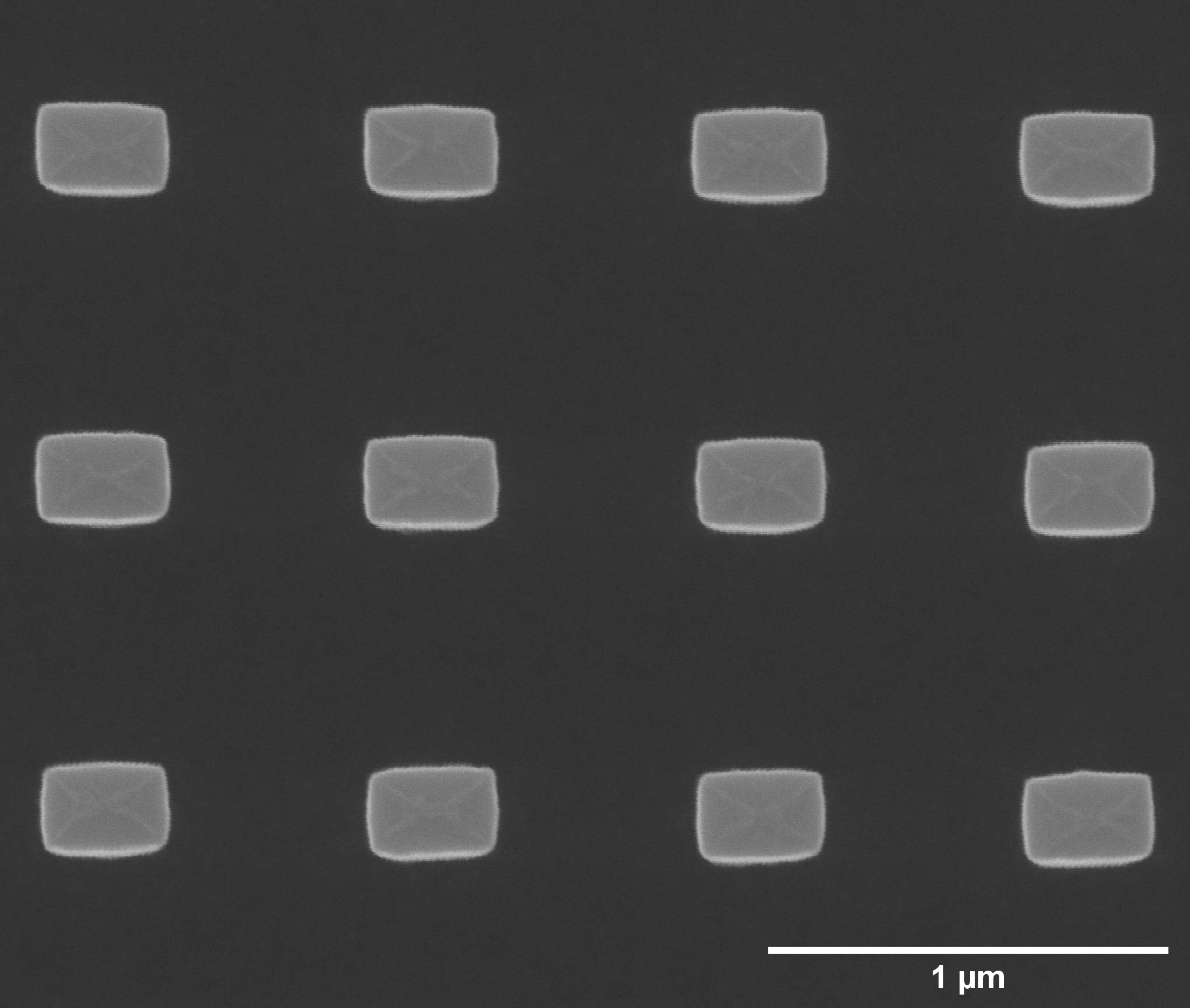}
         \caption{Top view}
         \label{fig:SEM_TopView}
     \end{subfigure}
        \caption{\textbf{Geometry of the metasurface}.(a)-(b) Illustration of a geometric phase (GP) metalens seen from front (a) and from rear side (b). The front and rear view are interchanged by flipping the metalens around the y-axis (i.e. the rotation axis). Doing so is seen to correspond to a sign change in the rotation angle $\alpha \to -\alpha$ of the rectangles. Hence, if the rotation angles $\alpha(r)$ corresponding with various radii $r$ are chosen so as to implement the phase function of a lens $\theta(r)=2\alpha(r)$, then the one side will correspond to a positive (converging) lens whereas the other will correspond to a negative (divergent) lens for a given circular polarization state. (c)-(d) SEM images of fabricated Silicon metasurface by UV-Nanoimprint lithography and Bosch DRIE etching. The details of the fabrication process are discussed in Sec. \ref{sec:MetasurfaceFabrication} and more details are given in \cite{dirdal2023uv}.}
        \label{fig:GSPMetalenses}
\end{figure}

The direction dependent focusing/defocusing of the GP-metalens shall now be explained. Geometric phase (also known as Pancharatnam-Berry phase) allows for pointwise implementation of a lens phase function by rotated structures, working on circular polarization states of light \cite{kang2012wave, khorasaninejad2016metalenses}. The operation $\hat{T}$ of an array of identical metasurface elements on right $|R\rangle$ and left $|L\rangle$ circular polarized light can be shown to place the transmitted light in a superposition of circular polarization states according to

\begin{eqnarray}
\hat{T}|R\rangle &=& B \exp(i2\alpha)|L\rangle + A|R\rangle, \label{eq:RPol} \\
\hat{T}|L\rangle &=& A |L\rangle + B \exp(-i2\alpha)|R\rangle. \label{eq:LPol}
\end{eqnarray} As is observed, the resulting cross-polarized field has attained a phase $\pm2\alpha$, where $\alpha$ can be shown to be equal the rotation angle of the metasurface inclusions: For our implementation the inclusions are rectangular pillars (Fig. \ref{fig:TwoReflectiveLenses}). Full cross-polarization ($|A|^2 \to 0$ and $|B|^2 \to 1$) can be achieved through suitable design of the metalens elements \cite{dirdal2023uv}. To implement a positive (converging) lens phase function $\theta(r)$ which varies circular-symmetrically with radius $r$ over the metalens, the rotations of the rectangular pillars are varied pointwise according to $\alpha(r)=\theta(r)/2$ in the layout. Note that the resulting metasurface is thus not an array of identical elements, and Eqs. \eqref{eq:RPol} and \eqref{eq:LPol} are therefore considered approximations. Figures \ref{fig:IncomingMetasurface} and \ref{fig:OutgoingMetasurface} illustrate the front and rear view of a GP-metalens. It may be observed that flipping between front and rear face of the metasurface is equivalent to a sign change in the rotation angles of the rectangular pillars $\alpha (r) \to -\alpha(r)$, as indicated in the two zoom-in portions of the metalens displayed (red solid and blue dashed squares). With a sign change on all angles, the reflected metalens thus implements a lens function of opposite sign $\theta(r)\to - \theta(r)$. Therefore, assuming the front side of the metalens implements a positive (converging) lens, the opposite side therefore implements a negative (diverging) lens, or vice versa. 

Given the observation above, it is natural to ask whether flipping a geometric phase metasurface always is equivalent changing the sign of the corresponding rotation angles. It turns out that this is only the case when the sub-unit structures are (i) individually reflection symmetric, and (ii) the rotations of each equidistant pair of sub-units from the reflection axis are identical. This is explained in further detail in Sec. \ref{sec:Discussion}. As a final remark on the metalens design, note that the above discussion presumes that the incidence light has the same circular polarization state traveling through the front and rear of the metalens. Recalling from Eqs. \eqref{eq:RPol}-\eqref{eq:LPol} that the light is cross-polarized after transmission, our compact-reflective lens (CRM) concept relies on the mirror to ensure that the light has the same circular polarization state upon transmission through the GP-metalens in both directions: After first transmission through the GP-metalens, the mirror reverts the polarization state from the cross-polarized light (e.g. $|L\rangle$) back to the incident polarization state (i.e. $|R\rangle$), before the light re-enters the GP-metalens the second time.

\subsection{Thin lens model of the Compact Reflective Lens} \label{sec:ThinLensModel}
\begin{figure}[h]
     \centering
     \begin{subfigure}[b]{0.4\textwidth}
         \centering
         \includegraphics[width=\textwidth]{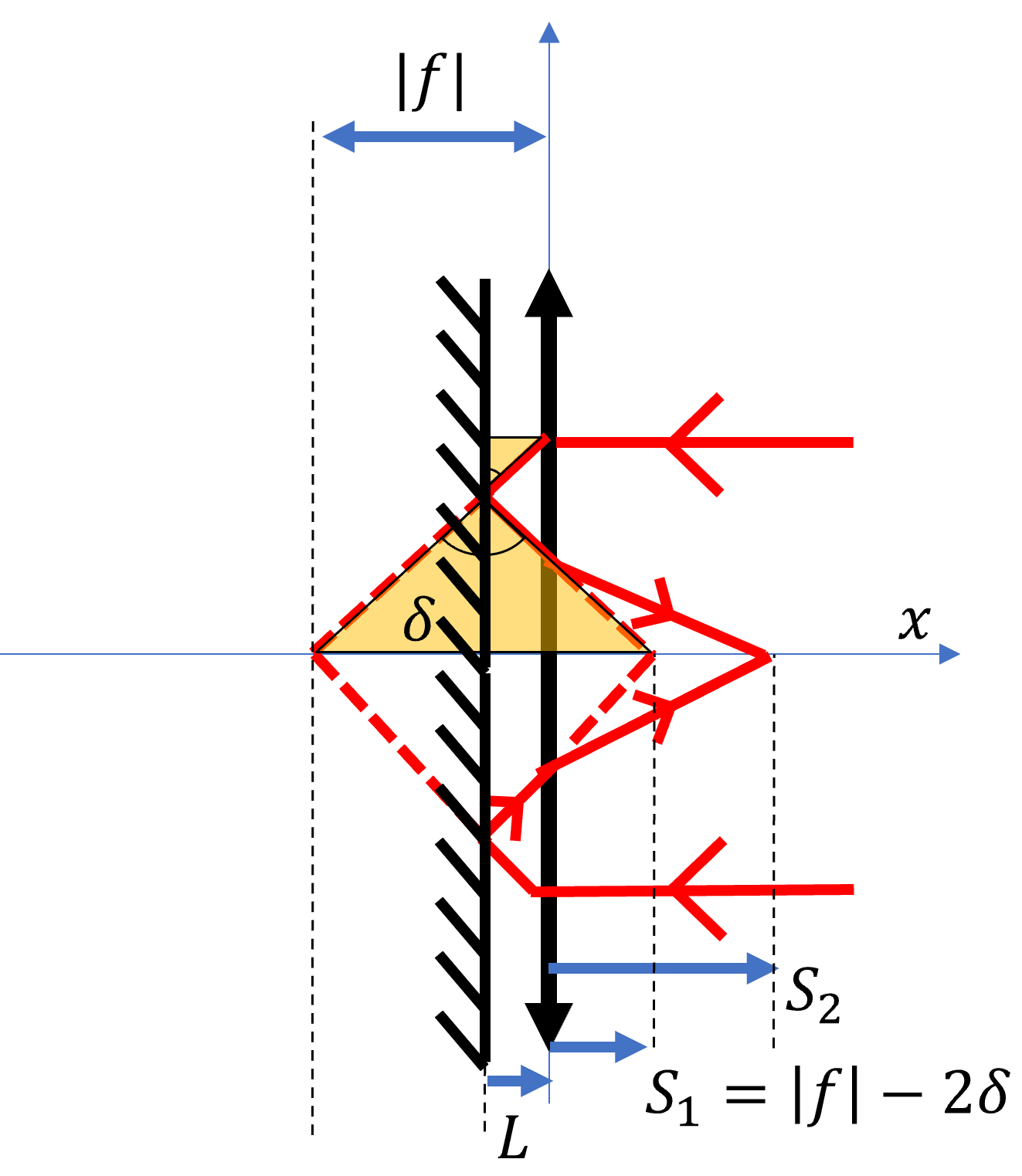}
         \caption{}
         \label{fig:ThinLensModel}
     \end{subfigure}
     \hfill 
     \begin{subfigure}[b]{0.5\textwidth}
         \centering
         \includegraphics[width=\textwidth]{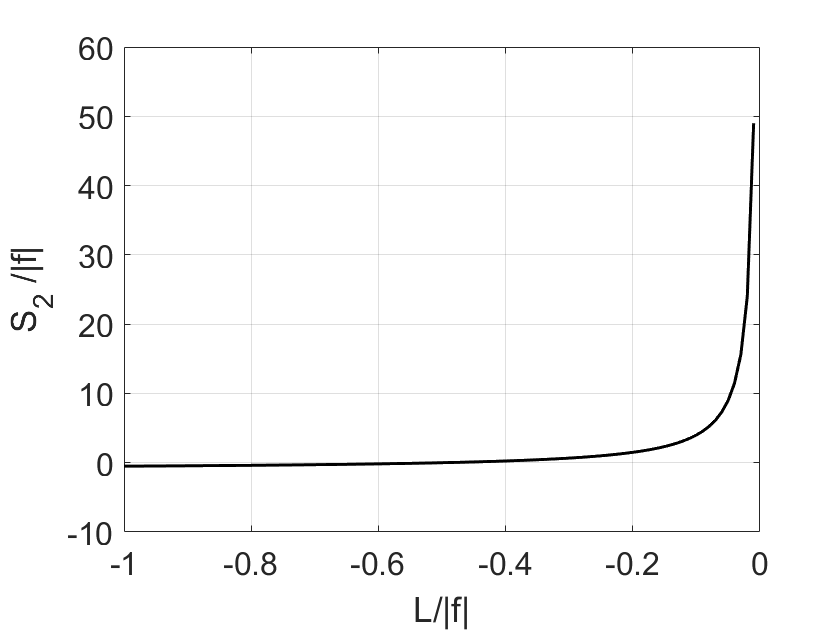}
         \caption{}
         \label{fig:NormalizedPlot}
     \end{subfigure}
        \caption{(a) Same compact reflective lens (CRM) as illustrated in Fig. \ref{fig:MetaReflectiveLens} with defined quantities and ray-tracing for deriving the thin lens model in Eqs. \eqref{eq:S2}. (b) A normalized plot of effective focal lengths from GP-metalens of the CRM upon normalized lens-mirror displacement by the thin-lens model when assuming $|R\rangle$.} 
        \label{fig:Model}
\end{figure}

This section presents a thin-lens model of the CRM. Despite the simplicity of the model it coincides well with both experimental measurements and Zemax simulations (see Sec. \ref{sec:Results} and Supplementary Information 2.1). In Fig. \ref{fig:Model} relevant parameters are defined for deriving the effective focal point distance $S_2$ of the reflective lens in terms of the lens-mirror spacing $L$, assuming collimated incoming light and a GP-metalens designed to be positive upon first transmission for right circular polarization $|R\rangle$ and negative for second transmission of $R\rangle$ (after reflection from the mirror). The dashed lines identify the refracted rays directed towards the lens focal point at a distance $|f|$ from the GP-metalens. A triangle with base length $\delta$ is drawn to help express the distance to the object point $S_1$ for second (reflected) transmission through the lens. Upon second transmission the GP-metalen will act as a negative lens with focal length $-|f|$, for which application of the thin lens equation gives the final focal point distance from the GP-metalens:

\begin{eqnarray}
    S_2 &=& -\frac{|f|(|f|-2\delta)}{2|f|-2\delta}, \nonumber \\
    &=& -\frac{|f|(2L+|f|)}{2L}. \label{eq:S2}
\end{eqnarray} The lens-mirror spacing is given by $L=\delta - |f|$ having set the origin at the lens intersection with the optical axis (i.e. assuming all lens-mirror spacings are negative $L=-|L|$). The solid curve in Fig. \ref{fig:NormalizedPlot} plots the normalized effective focal point $S_2/|f|$ vs the normalized lens-mirror spacing $L/|f|$. As expected one observes rapid change in the focal length $S_2$ close to zero gap $L=0$, and the effective focal point approaches $S_2 \to -|f|$ at large gaps (as expected, corresponding with the image point for a negative lens when an object is placed at infinity). At $L=-0.25|f|$ the effective focal length becomes equal to the focal length of the metalens $S_2= |f|$. A MEMS-displacement range of 25\% of the focal length of the lens $|f|$ therefore allows to modulate the focal point between infinity and a distance $|f|$ from the metalens.

Note that it is possible to combine the metalens with a refractive surface beneath the metasurface structures or on the back side of the silicon substrate. This allows also for shifting the curve of Fig. \ref{fig:NormalizedPlot} so as to achieve a finite focal length at zero lens-mirror gap. This design freedom is further discussed in the SI 1.1.

The preceding discussion assumes a metalens designed to act as a positive lens upon first transmission for right circular incoming light $|R\rangle$. In the SI 1.2 it is shown that applying cross-polarized light (i.e. $|L\rangle$) to CRM, the device becomes equivalent with a doublet consisting of consecutive negative and positive lenses with equal focal length magnitudes.

\section{Implementation}

\subsection{Fabrication of metasurface lens} \label{sec:MetasurfaceFabrication}
The metasurface design employed in this publications targets an operation wavelength of 1550nm and is fabricated in silicon by use of UV-nanoimprint lithography (UV-NIL) patterning of resist and subsequent Bosch Deep Reactive Ion Etching (DRIE). The resutling structures are shown in Fig. \ref{fig:SEM_Crossection} and \ref{fig:SEM_TopView}. Two metalenses were utilized, one with focal length $|f|=10$ mm and size 1.5 mm x 1.5 mm, and one with $|f|\approx 215$ $\mu$m with size 300µm x 300µm. The latter lens does not have a well-defined focal length because its phase function was found through Zemax optimization of the lens placed in the doublet configuration utilized in this paper. As mentioned,  the phase function is implemented in terms of pointwise rotations of the silicon pillars at the center of each unit cell. The targeted rectangular pillar dimensions of the UV-NIL template are: Widths $w=237$nm, lengths $l=361$nm, heights $h=1200$nm, and the pitches of the square unit cells are $p=835$nm.

As can be seen from Fig. \ref{fig:SEM_Crossection} and \ref{fig:SEM_TopView}, smooth sidewalls are attained despite using the three steps passivation, de-passivation and isotropic etch of the Bosch process: The Bosch process was tuned to aim for a lateral etch depth (scallop depth) on the order of 10nm. A detailed account of this process offering structures with good pattern fidelity and high focusing efficiency have been earlier published \cite{dirdal2023uv}.

\subsection{Novel micromirror with long stroke piezoelectric MEMS actuation} \label{Sec:MEMS}

\begin{figure}[h]
    \centering
    \includegraphics[width=0.9\textwidth]
    {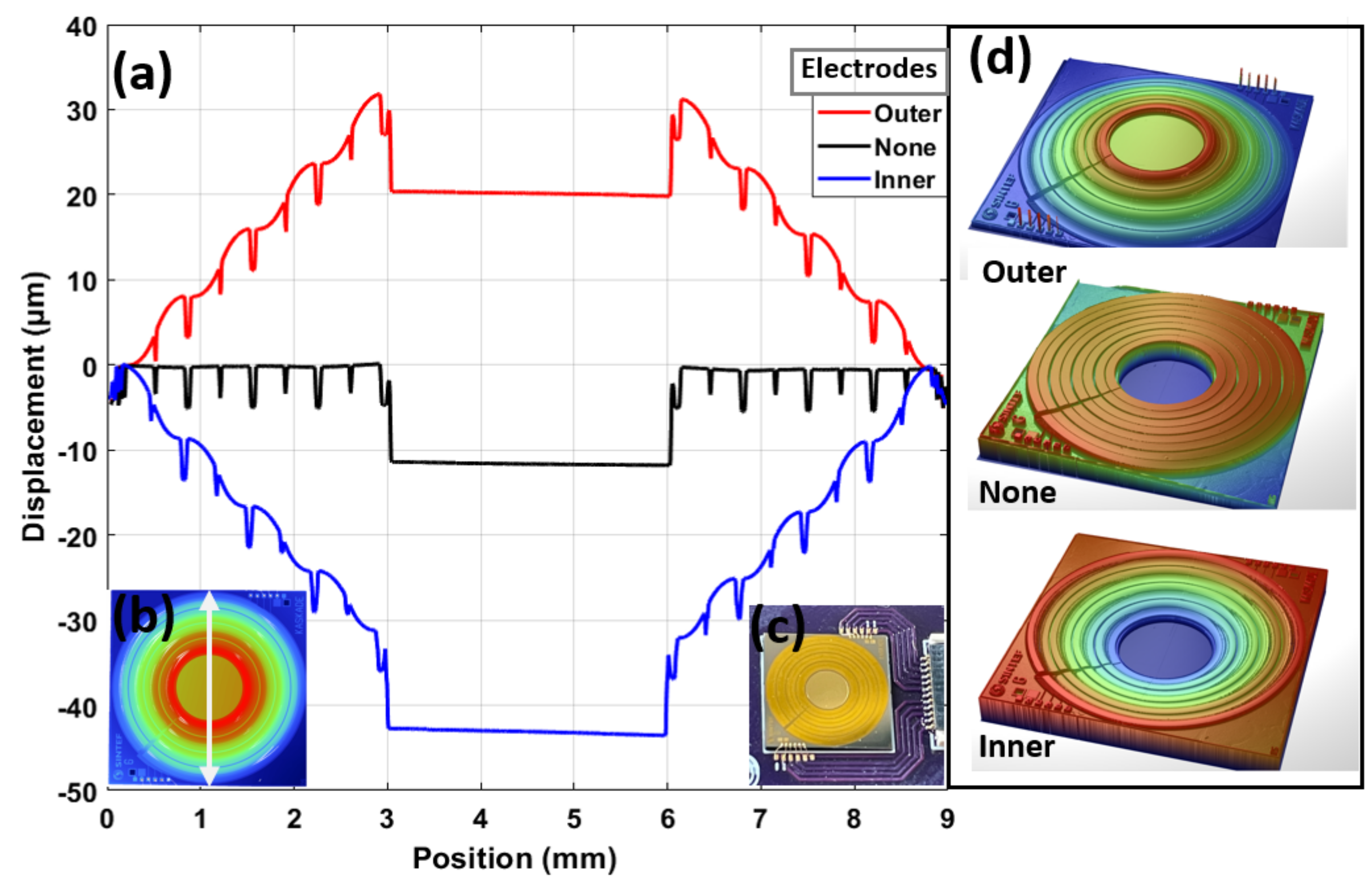}
    \caption{\textbf{Long-stroke MEMS-micromirror.} (a) Cross-sectional height profile of the MEMS micromirror along the line indicated in the inset (b), as measured by white-light interferometry. The mirror height corresponding to three distinct actuation states is shown. At a 40V application to the innermost electrodes, the PZT membrane pulls the ring downward, placing the mirror at its lowest position. Conversely, a 40V application to the outermost electrodes causes the PZT membrane to draw the ring upward, resulting in the highest position. In the absence of voltage, an intermediate position is achieved. Notably, a displacement range of 62 µm is observed, and a high level of planarity is maintained throughout movement. (c) Picture of a MEMS mirror. (d) Presents 3D images of the MEMS mirror at the three specified actuation states.}
    \label{fig:Whitelight-MEMS}
\end{figure}

We have previously demonstrated a piezo-electric actuated micromirror which can achieve tilting motion of ±0.5° degrees and translational "piston" motion of 7.2µm \cite{bakke2010novel, Dirdal2022mems}. A suspended central circular gold mirror of diameter 3mm is actuated by applying voltages to a thin-film PZT piezoelectric membrane encircling it, deposited on typically 700µm wide and 10µm thin actuator arms. A similar concept is realized here, using a cascade of concentric actuator-rings to increase the stroke-length. Here, the downward/upward push/pull of diameter 3 mm, 400 µm thick, planar gold-plated central mirror is generated by actuating four independent segments of thin film PZT actuator-pairs. As shown in Fig. \ref{fig:Whitelight-MEMS}, biasing the inner actuators relative to the mirror, push the mirror down, while biasing the outer rings pull it up. It is foudn that 10 µm thick membrane sections of 700 µm width maximize the piston stroke-length while maintaining mirror-planarity and stability, given the mechanical boundary-conditions for the present design.

A vertical displacement of up to ± 31 µm (a 62 µm total stroke) at 40V of applied voltage has been measured using white light interferometry, as shown in Fig. \ref{fig:Whitelight-MEMS}. This is an order of magnitude larger than that of our previously reported micromirror \cite{bakke2010novel, Dirdal2022mems}. Versions in which the electrode segments have been further divided into quadrants allow for enabling tip/tilt motion, e.g. for deflecting a laser beam. Observing a reflected laser beam under tip/tilt actuation from 0.5 kHz to 6 kHz reveal resonances at around 1.86 kHz, 2.8 kHz and 4.5-5.5kHz.

Thin-film piezoelectric MEMS has the advantage of requiring ultra low power. From similar architectures previously characterized we have measured power consumption of a few 100nW at 23V. 
Thin-film piezoelectric PZT MEMS can offer long-term durability and reliability making them relevant for wide range of applications\cite{dahl2020effect, dahl2022reliable}.

\section{Results}\label{sec:Results}

\subsection{Verification of lens concept} \label{sec:Verification}
\begin{figure}[h]
     \centering
     \begin{subfigure}[b]{0.5\textwidth}
         \centering
         \includegraphics[width=\textwidth]{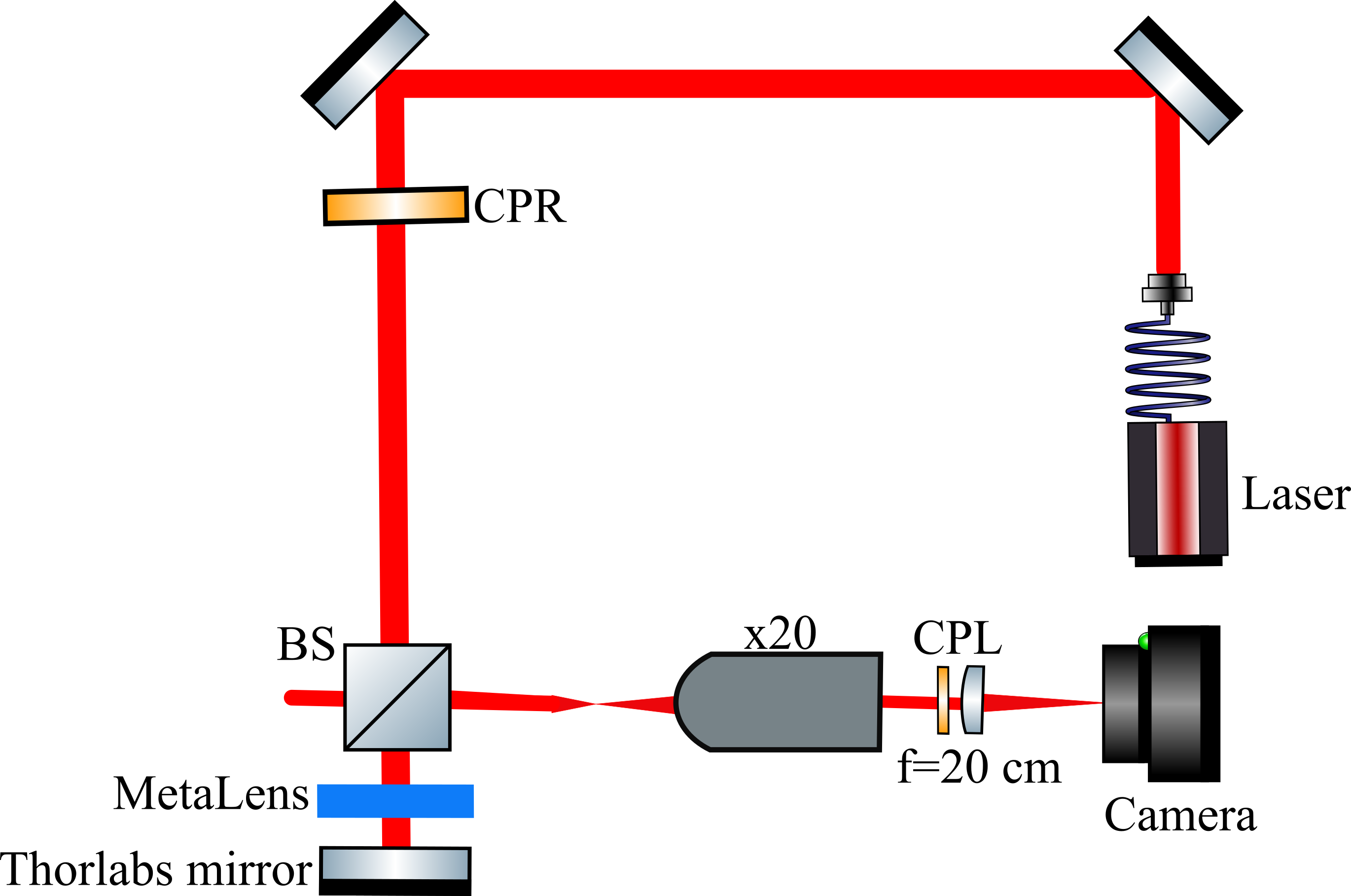}
         \caption{Experimental setup}
         \label{fig:ExperimentalSetup}
     \end{subfigure}
     \hfill 
     \begin{subfigure}[b]{0.45\textwidth}
         \centering
         \includegraphics[width=\textwidth]{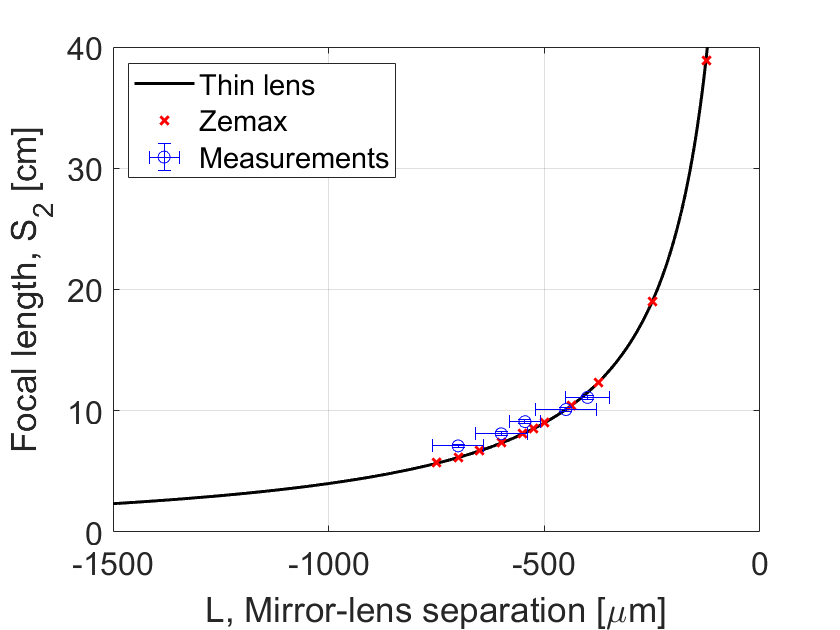}
         \caption{Prediction based on thin lens model and Zemax simulations, and experimental results}
         \label{fig:PlottedResults}
     \end{subfigure}
    \begin{subfigure}[b]{0.4\textwidth}
         \centering
         \includegraphics[width=\textwidth]{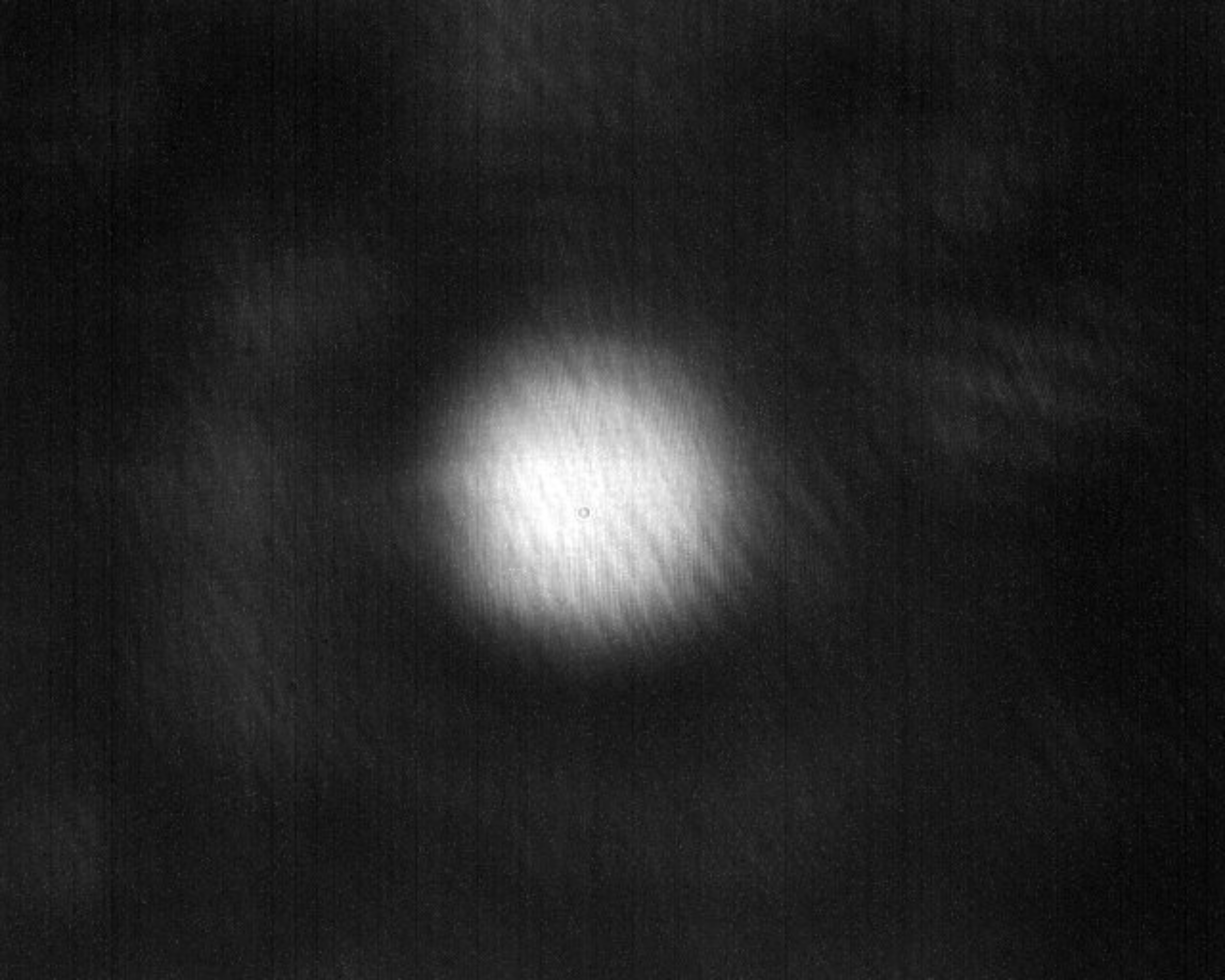}
         \caption{Focus}
         \label{fig:FocusedImage}
     \end{subfigure}
    \begin{subfigure}[b]{0.4\textwidth}
         \centering
         \includegraphics[width=\textwidth]{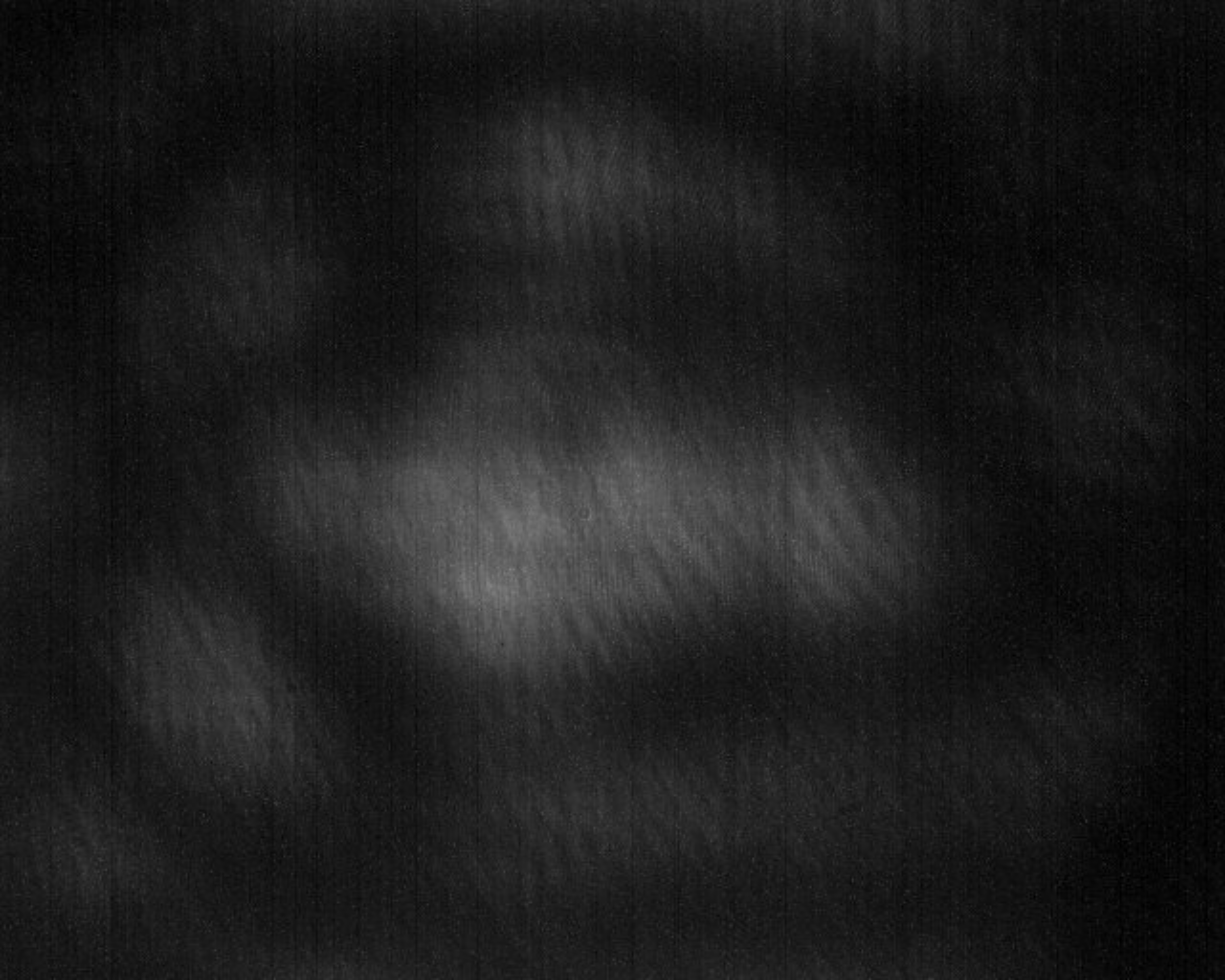}
         \caption{Defocus}
         \label{fig:DefocusedImage}
     \end{subfigure}
        \caption{\textbf{Tuning the effective focal length using reflective lens.} (a) Schematic of the optical setup.
        (b) Outcome from the thin lens model, Zemax simulations and experimental results. Focus (c) and defocus (d) of the reflective lens by changing the lens-mirror displacement.}
        \label{fig:10mmLens}
\end{figure}

\begin{figure}[h]
\centering
     \begin{subfigure}[b]{0.525\textwidth}
         \centering
         \includegraphics[width=\textwidth]{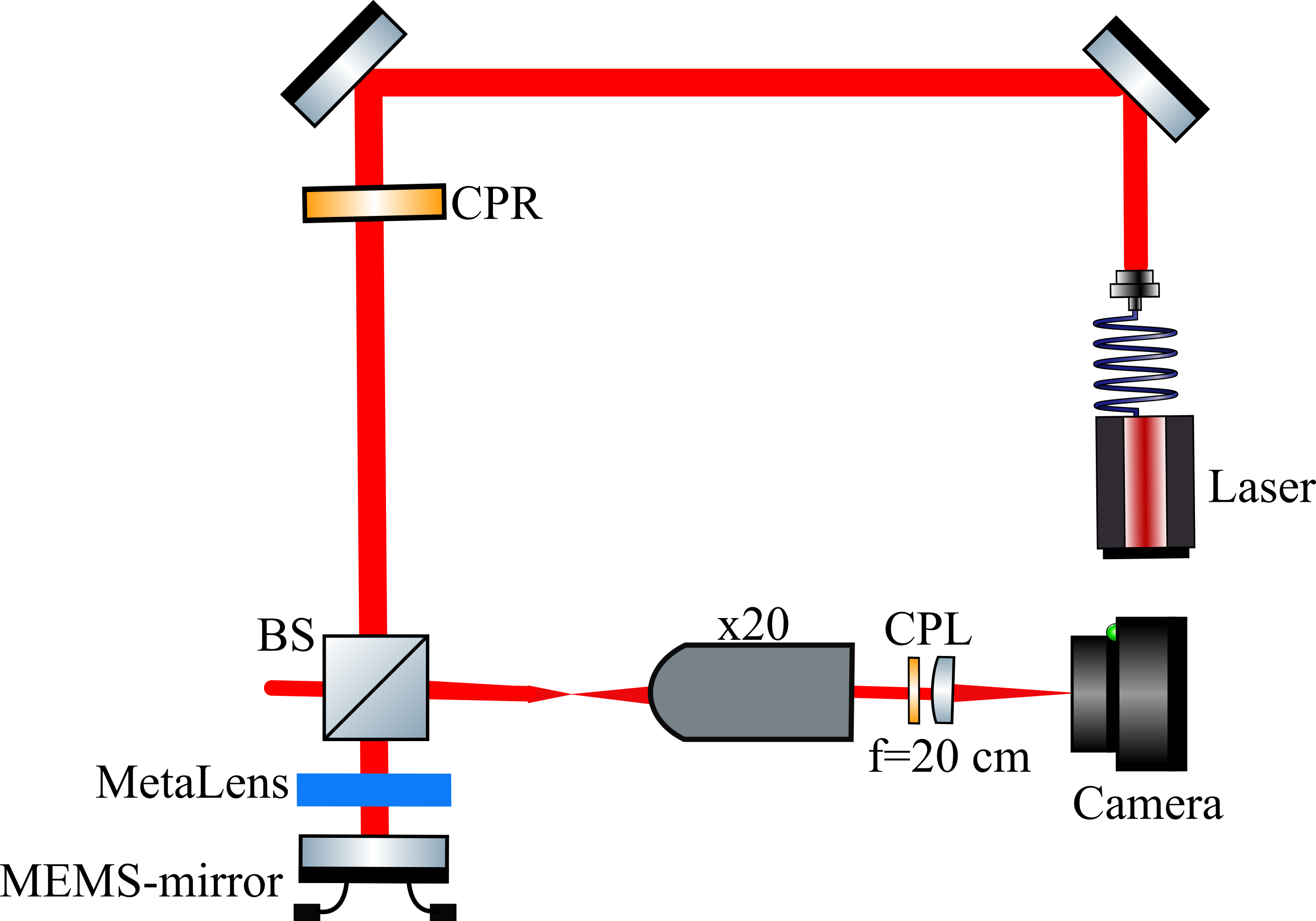}
         \caption{Experimental setup}
         \label{fig:ExperimentalSetup2}
     \end{subfigure}
     \hfill 
       \centering
    \begin{subfigure}[b]{0.45\textwidth}
         \centering
         \includegraphics[width=\textwidth]{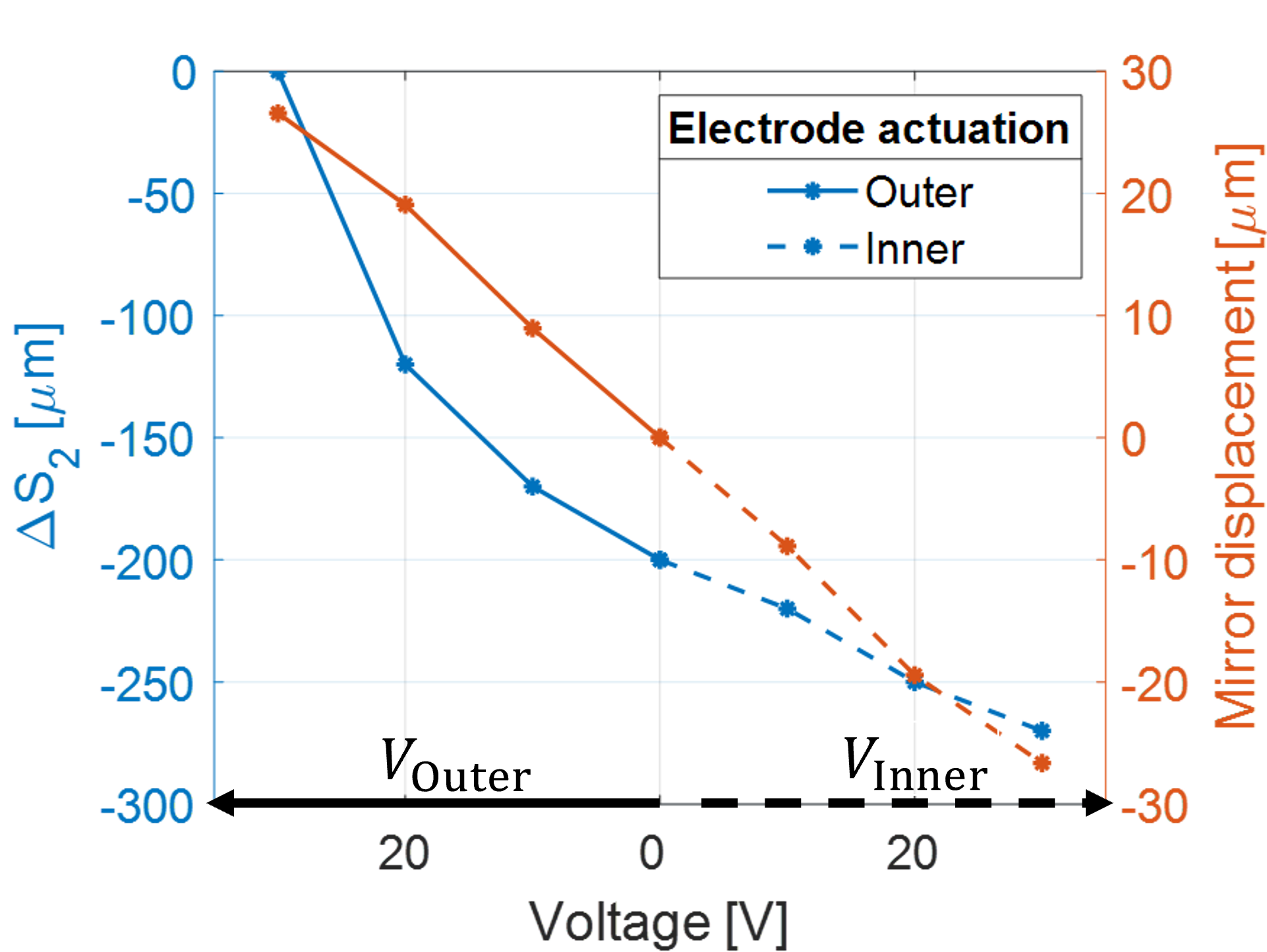}
         \caption{MEMS actuation}
         \label{fig:S2vsVoltage2}
     \end{subfigure}
     \hfill 
     \centering
     \begin{subfigure}[b]{0.45\textwidth}
         \centering
         \includegraphics[width=\textwidth]{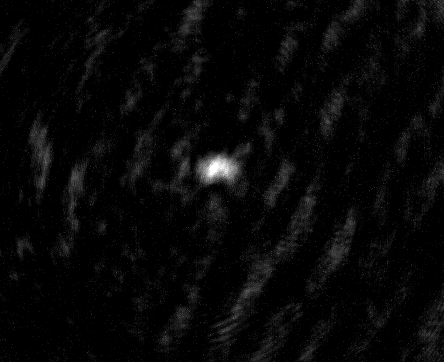}
         \caption{Focus}
         \label{fig:FocusMEMS2}
     \end{subfigure}
     \hfill 
     \begin{subfigure}[b]{0.45\textwidth}
         \centering
         \includegraphics[width=\textwidth]{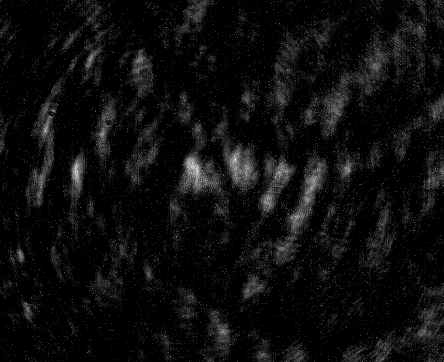}
         \caption{Defocus}
         \label{fig:DefocusMEMS2}
     \end{subfigure}
    \caption{\textbf{Tuning the effective focal length using CRM.} (a) Schematic of the optical setup. (b) Left axis: Experimentally measured focal length shift vs voltage actuation. Right axis: Experimentally measured mirror displacement vs voltage application. Focus (c) and defocus (d) of the CRM by changing the lens- MEMS mirror displacement.}
    \label{fig:MEMS-tunningl}
\end{figure}

To verify the reflective metalens concept, a mirror was displaced from the GP-metalens using the setup illustrated in Fig. \ref{fig:ExperimentalSetup}. First manual displacement of the mirror was used with a metalens of focal length $|f|=10$mm and the results are compared with Eq. \eqref{eq:S2} of the thin lens model, and thereafter the MEMS micromirror presented in Sec. \ref{Sec:MEMS} was used together with a metalens of focal length $|f|\approx 215\mu$m to verify the compact refletive metalens (CRM) concept using the setup illustrated in Fig. \ref{fig:ExperimentalSetup2}.

As shown in Fig. \ref{fig:ExperimentalSetup}, the lens-mirror doublet is illuminated by a collimated laser beam from a 1.55µm fiber laser, which undergoes dual reflection off two mirrors (for alignment). It then passes through a right-handed circular polariser (CPR) to place it in right circular $|R\rangle$ state. A 50/50 non-polarizing beam splitter (BS) is used for illuminating and collecting along normal incidence angle. The light then travels through the GP-metalens, is cross polarized to $|L\rangle$ and focused. The light then reflects off the mirror positioned behind the GP-metalens, reverting the polarization state from $|L\rangle$ to $|R\rangle$ again. The reflected light re-enters the GP-metalens a second time, is cross-polarized to $|L\rangle$ and then diverged. The light then passes through the beam splitter. The light for imaging is directed through a x20 infinity-corrected microscope objective, a tube lens, and an IR-camera. In reality, due to fabrication imperfections, not all light is cross-polarized during the transmissions through the metalens. Improved image quality is achieved by selecting only the $|L\rangle$ light, by use of a left-handed circular polarizer (CPL) placed between the microscope objective and the planoconvex lens. Both the metalens and mirror are mounted to an XYZ translation stage for alignment and distance adjustments. Additionally, the mirror is attached to a kinematic mount, allowing for a few degrees of tip and tilt adjustment. The objective lens is mounted on a single-axis motorized translation stage equipped with a controller for precise movements (with step accuracy $1\mu$m).

Figures \ref{fig:FocusedImage} and \ref{fig:DefocusedImage} show focusing and defocusing of the reflective lens by changing the lens-mirror displacement. When the metalens and mirror are well-aligned and parallel a strong back reflection (from the air-silicon interface) occurs which obstructs the focused spot. However, by carefully adding a slight mirror tilt, we successfully separated the focus spot and selectively bringing it into the camera's field of view and thus significantly reduce interference from strong back reflections. Addition of an anti-reflectance coating would be helpful in reducing such back reflections from the silicon substrate. To compare the the performance of the reflective lens with the thin lens model of Eq. \eqref{eq:S2}, an experiment was performed where the focal distance was found for various lens-mirror displacements and plotted in Fig. \ref{fig:PlottedResults}: For each position of the objective lens, the focal point of the reflective metalens is brought into view by modulating the lens-mirror distance. The focal length $S_2$ is calculated by first finding the focal point with the objective lens, and then subtracting the working distance of the objective lens (20mm) from the measured distance between the silicon backside of the metalens to the objective lens. The objective lens is moved using the motorized translation stage. The lens-mirror separation $L$ is found from first bringing the metalens and mirror into contact $L\to 0$ and then noting the rotation of the micrometer screw which translates the mirror, as the lens-mirror gap is increased. The minimum unit of the micrometer screw is 10µm. After finding the focal spot with the objective lens for each measurement, slight adjustments to the mirror position which leads to visible defocus on either side of the focus are used to define the measurement uncertainty, and the average value is used to calculate the actual lens-mirror spacing. The resulting measurements are represented as blue dots (with error bars) in Fig. \ref{fig:PlottedResults}. For the metalens with focal length $|f|=10$ mm a focal tunability of 4 cm was measured by changing the separation between the mirror and metalens from 400 µm to 700 µm. Notably, the measurement results closely align with the predictions of the thin lens model given by Eq. \eqref{eq:S2}. The thin lens model is furthermore verified with results obtained from simulating the system in Zemax (red crosses) - more information on this is given in SI 2.1. 

To verify a compact reflective metalens (CRM) we used the long stroke displacement MEMS mirror of Sec. \ref{Sec:MEMS} which offers 62 µm for an applied voltage of 40V, as shown in Fig. \ref{fig:Whitelight-MEMS}. To ensure large diopter changes, a focal length is chosen that corresponds with the available MEMS range (as discussed in Sec. \ref{sec:ThinLensModel}): We combined the micromirror with the metalens of focal length $|f|\approx 215\mu$m (Sec. \ref{sec:MetasurfaceFabrication}). According to the normalized, thin-lens model plot of Fig. \ref{fig:NormalizedPlot}, this MEMS-range should in theory allow for a considerable diopter change of around 6330 m$^{-1}$ by shifting the focal point from infinity to 157$\mu$m away from the metalens. Figures \ref{fig:FocusMEMS2} and \ref{fig:DefocusMEMS2} show focusing and defocusing of the CRM by changing the lens-MEMS mirror displacement. In Fig. \ref{fig:S2vsVoltage2} we present the change in focal length vs applied voltage to the MEMS-actuated CRM. For the $|f|\approx 215\mu$m lens, the uncertainty of measurement of both focal length and lens-mirror gap are considerable compared to the involved length scales. At the starting point, where a voltage $V_\text{Outer}=30$V is applied to the outer electrodes of the micromirror, neither lens-mirror gap or focal distance are therefore known with certainty. However, given the precise movement ($1\mu$m) of the objective lens due to movement by the motorized stage, accurate measurements of the focal length \emph{change} corresponding to voltage application to the outer and inner electrodes of the micromirror are obtained and presented in Fig. \ref{fig:S2vsVoltage2}. Furthermore, micromirror displacements corresponding to voltage application are known from separate white-light interferometric measurements (Fig. \ref{fig:Whitelight-MEMS}), and are added to the right y-axis of Fig. \ref{fig:S2vsVoltage2}. Combined, the experimental results demonstrate a focal shift of $\Delta f_\text{meas}=270\mu$m for a micromirror displacement of $\Delta L=53\mu$m from a actuation voltage of 30V between outer and inner electrodes.

An estimate of the corresponding diopter change to the focal length change can be obtained by fitting the measured focal length change $\Delta f_\text{meas}$ and micromirror shift $\Delta L$ to the thin-lens model curve, when accounting for the silicon substrate. Under paraxial assumptions the correction factor $t(1-n_\text{air}/n_\text{Si})$ should be added to the thin lens model focal length $S_2$, where $t=500\mu$m and $n_\text{Si}=3.5$ are the thickness and refractive index of the silicon substrate, respectively, and $n_\text{air}=1$ is the refractive index of air. This fitting leads to estimated start and end focal lengths of $f_1=650\mu$m and $f_2=376\mu$m, respectively, and corresponding lens-mirror gaps of $L_1=-45.5\mu$m and $L_2= -98.5\mu$m (see SI 2.2 for more details). From these values the diopter change is estimated to be 1121 m$^{-1}$.

Given the substrate thickness $t=500\mu$m, the estimate $f_2=376\mu$m indicates that the focal point of the CRM has entered into the silicon substrate where both the focal point and the working distance of the objective lens become shifted relative to their position in air due to refraction. Under paraxial assumptions it is easy to show that the effects of the focal point and working distance shifts compensate, so that the measured focal distance under the setup described above is equivalent to measuring the focal distance outside of the silicon (see SI 2.2 where also Zemax simulations are shown to confirm this).

\section{Discussion} \label{sec:Discussion}

An interesting property of the geometric phase metalens is that flipping it around its central axis is equivalent to changing the signs on the rectangle rotations, as discussed in Sec. \ref{sec:GPM_makes_CRM}. As mentioned there, this is not a general result applicable to all geometric phase metasurfaces. The specific conditions shall now be discussed. Mathematically, flipping the metalens around the central axis (about the y axis in Fig. \ref{fig:OutgoingMetasurface}) corresponds to a reflection transform of the metalens. For a shape to equal its reflection transform apart from differences in rotations, it must itself be reflection symmetric. This can be shown by demanding that the sets of points of a shape and its reflection transform coincide when the shapes are suitably rotated and translated to the origin (SI 3). As a contrary example, one may for instance draw right angled triangles between three corners of the shown rectangles (i.e. non-reflection-symmetric sub-units) displayed in Fig. \ref{fig:IncomingMetasurface} and observe how the reflection transform no longer corresponds to a sign change in rotations. In addition to having reflection symmetric structures, it matters how they are collectively placed: There must be identical placement of structures on either side of the rotation axis. In Fig. \ref{fig:OutgoingMetasurface} one can observe pairwise identical rectangles placed on either side of the central column of rectangles in every row. This ensures that the reflection transformed sub-units are placed at the coordinates that correspond to their oppositely rotated counterparts in the non-reflected metasurface. This requirement is automatically fulfilled when implementing the phase function of a lens $\theta(r)$ because of its circular symmetry. As a contrary example, if one considers the reflection transform of the zoom in portion of Fig. \ref{fig:IncomingMetasurface} (enclosed by a red solid square) placed over itself, it may be observed in some of the rectangles that the transformation is not equivalent to a sign change in rotation. This occurs due to the lack of pairwise identical placement of rectangles about the central column of the zoom-in portion. As a side note, observe that due to the inversion symmetry of the rectangular structures used in this work (in addition to their reflection symmetry), the metasurface in Fig. \ref{fig:IncomingMetasurface} attains inversion symmetry.

\section{Conclusion}
Throughout the history of optics, lenses have been either converging or diverging. In this work we present a notable exception to this rule, using a geometric phase metasurface lens (metalens) which can both act as a positive (converging) or negative (diverging) lens, depending on transmission through the front or rear of the lens, respectively. This is enabled by the geometric phase principle and certain symmetry requirements outlined in this work. The unique property of the metalens allows for certain freedom in design of compound lenses. In this work we present a varifocal reflective doublet concept that is equivalent to a transmissive doublet with a consecutive positive and negative lens. This enables an ultra-compact tunable lens of large diopter changes, which is only possible with the given metalens. The lens works without the need for a spacer between the metalens and mirror, making the device well-suited for wafer-level silicon fabrication at high volumes and low cost. 

A proof-of-concept implementation using a 1550nm NIR metalens is demonstrated, coupled with a novel long stroke (62 µm at 40V) and high resonance frequency ($\sim 1.86$ kHz) piezoelectric MEMS-micromirror. The metasurface displacement range achieved by this MEMS-mirror surpasses that of state-of-the-art electrostatic actuation by more than an order of magnitude, and theoretically enables a considerable diopter change on the order of 6330 m$^{-1}$. Our proof-of-concept implementation demonstrated a diopter change of around 1121 m$^{-1}$. The manuscript also presents a thin lens model of the compact reflective lens (CRM) and explains the geometric phase metalens working principle. Additionally, the manuscript provides a detailed account of the design and working principle of the novel long-stroke MEMS-micromirror. 

Experimental results validate the CRM concept, demonstrating focal tunability through manual adjustments of the lens-mirror separation in the case of a 10mm focal length metalens, and voltage-controlled displacement of a MEMS mirror configuration in the case of a 215µm metalens. The experimental outcomes closely align with simulations and the derived thin lens model. 

Overall, the proposed reflective lens system offers a promising solution for achieving fast and strong modulations in a ultra-compact form factor. The large focal length shifts and diopter changes it enables is of relevance to a wide range of micro-optical applications within sensing and imaging.

\backmatter

\bmhead{Supplementary information}
Please refer to the Supplementary Information for further details.

\bmhead{Acknowledgments}

This research was funded by EEA Financial Mechanism 2014–2021, under project no. EEA-RO-NO2018-0438—ElastoMETA, and the ATTRACT programme that has received funding from the European Union's Horizon 2020 Research and Innovation Programme under Grant Agreement no. 101004462. We thank our colleagues Anand Summanwar and Karolina Milenko for assistance with fabricating the metalenses, Aina Herbjørnrød for assistance with fabricating the micromirrors.

\bmhead{Author Contributions}
CD conceived of the presented idea and developed the theory. GB performed the ray tracing simulations with input from CD. The design and fabrication of metasurface and MEMS micromirror were lead by CD and ZS, respectively. FD and JJ carried out the experiment with input from CD, GB and ZS. JJ performed the micromirror characterization. CD and FD wrote the manuscript with inputs from the other authors. All authors discussed the results and contributed to the final manuscript.

\section*{Declarations}
The authors declare no conflicts of interest.

\bibliography{sn-bibliography}

\end{document}